\documentclass[prl,twocolumn,superscriptaddress]{revtex4}

\usepackage{graphicx}
\usepackage{amsmath}
\usepackage{times}
\begin{document}

\title{Ultrafast insulator-to-metal phase transition as a switch to measure the spectrogram of a supercontinuum light pulse}

\author{Federico Cilento}
\affiliation{Department of Physics, Universit$\grave{a}$ degli Studi di Trieste, Trieste I-34127, Italy}
\author{Claudio Giannetti}
\affiliation{Department of Physics, Universit$\grave{a}$ Cattolica del Sacro Cuore, Brescia I-25121, Italy}
\author{Gabriele Ferrini}
\affiliation{Department of Physics, Universit$\grave{a}$ Cattolica del Sacro Cuore, Brescia I-25121, Italy}
\author{Stefano Dal Conte}
\affiliation{Department of Physics, Universit$\grave{a}$ degli Studi di Pavia, Italy}
\author{Tommaso Sala}
\affiliation{Department of Physics, Universit$\grave{a}$ Cattolica del Sacro Cuore, Brescia I-25121, Italy}
\author{Giacomo Coslovich}
\affiliation{Department of Physics, Universit$\grave{a}$ degli Studi di Trieste, Trieste I-34127, Italy}
\author{Matteo Rini}
\affiliation{Materials Sciences Division, Lawrence Berkeley National Laboratory, Berkeley, California, USA}
\author{Andrea Cavalleri}
\affiliation{Max Planck Research Group for Structural Dynamics, University of Hamburg, Germany}
\affiliation{Department of Physics, Clarendon Laboratory, University of Oxford, United Kingdom}
\author{Fulvio Parmigiani}
\affiliation{Department of Physics, Universit$\grave{a}$ degli Studi di Trieste, Trieste I-34127, Italy}
\affiliation{Elettra Synchrotron of Trieste, Basovizza I-34127, Italy}

\date{\today}

\begin{abstract}
In this letter we demonstrate the possibility to determine the temporal and spectral structure (spectrogram) of a complex light pulse exploiting the ultrafast switching character of a non-thermal photo-induced phase transition. As a proof, we use a VO$_{2}$ multi-film, undergoing an ultrafast insulator-to-metal phase transition when excited by femtosecond near-infrared laser pulses. The abrupt variation of the multi-film optical properties, over a broad infrared/visible frequency range, is exploited to determine, in-situ and in a simple way, the spectrogram of a supercontinuum pulse produced by a photonic crystal fiber. The determination of the structure of the pulse is mandatory to develop new pump-probe experiments with frequency resolution over a broad spectral range (700-1100 nm).\end{abstract}

\pacs{}

\maketitle

Recently, it has become possible to generate broadband continuum, extending from 430 to 1600 nm with a nearly flat spectral intensity, from few nanojoule pulses produced by a standard 120-fs 800-nm Ti:sapphire oscillator.
The key elements in continuum generation are the newly developed microstructured Photonic-Crystal Fibers (PCF), engineered to be nearly dispersion-free in the visible.
Non-linear interactions between an infrared laser pulse propagating into the fiber and the silica core generate a broadband non transform-limited pulse output with unknown residual spectral chirp \cite{Dudley2006}.
Since continuum is emerging as a new tool for time-resolved spectroscopic techniques with broadband frequency resolution \cite{Leonard2007}, it is mandatory to find simple methods to fully characterize the temporal and spectral structure (spectrogram) of complex light pulses.

Usually, non-linear gating techniques, such as Frequency-Resolved Optical Gating (FROG) \cite{Trebino:book}, are used to characterize continuum optical pulses \cite{Dudley2002}. These techniques require a phase matchable and transparent non-linear optical medium across the entire bandwidth of the pulse to be characterized. The problem of obtaining phase-matching conditions over a broad spectral range, with a single crystal, is generally solved by angle-dithering techniques \cite{Trebino:book}.

\begin{figure*}[t]
\centering
\includegraphics[keepaspectratio,bb= 40 375 580 610,clip,width=0.8\textwidth]{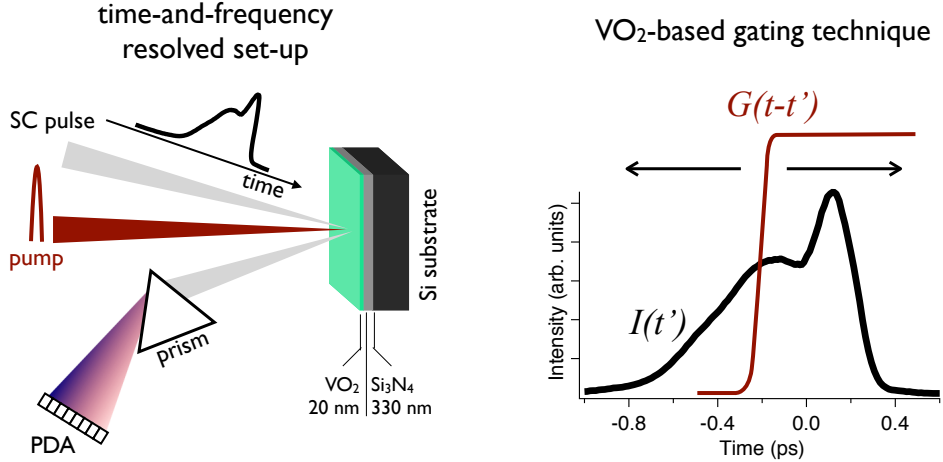}
\caption{\footnotesize{A sketch of the experimental set-up and of the VO$_2$-based gating technique is reported.}}
\label{fig1}
\end{figure*}

Although it is possible to use these non-linear techniques to fully characterize the continuum, practical problems in a typical pump and probe set-up (see Fig. \ref{fig1}) demands for an ad hoc solution \cite{Tien2009}.
The need to characterize the continuum pulses in situ, having access only to their reflection from a sample held inside a cryostat, possibly without the need of phase-matching conditions, prompted us to consider the possibility of a solid-state ultrafast optical switch as a gating element.
A thin VO$_{2}$ film, when excited with a near infrared ultrashort laser pulse, has the remarkable property of undergoing a photo-induced insulator-to-metal phase transition  \cite{Cavalleri2001,Cavalleri2004b}, resulting in an ultrafast change of the reflectivity over a broad frequency range.
In addition, the switching character of the system, due to the fact that the recovery time (10-100 ns) of the excited film is longer than the pulse duration, is also useful to check for multiple-pulses in the laser output.

In this letter, we demonstrate the application of a VO$_{2}$ thin-film multilayer sample as an ultrafast optical switch to perform a frequency and time characterization of a PCF-generated ultrabroad and ultrashort laser pulse, through a time resolved transient optical spectroscopy in a pump/probe configuration.
The major advantages of this technique are that no phase matching is needed, the fluence threshold for the process is reasonably low (\begin{math}\sim \end{math} 250  $\mu$J/cm$^{2}$ \cite{Rini2008}) to achieve, and the multilayer can be easily placed at the sample position without altering the optical design of the system, arranged to work in a reflectivity configuration.

A broadband electromagnetic field pulse with a complex temporal structure, i.e. $E(t)$=$[I(t)]^{1/2}{e}^{i\phi(t)}$=$\int{[\tilde{I}(\omega)]^{1/2}\mathrm{e}^{i\Phi(\omega)}\mathrm{e}^{i\omega t}d\omega/2\pi}$ ($I(t)$ being the slowly-varying intensity envelope), is characterized by the variation of the frequency with time (chirp) $\Omega(t)$=$-d\phi(t)/dt$ or, in the frequency domain, by $T(\omega)$=$-d\Phi(\omega)/d\omega$, i.e. the group delay of a particular slice of the spectrum at frequency $\omega$. The complete reconstruction of the time-frequency characteristics of the light pulse is called spectrogram and it is mathematically given by:
\begin{equation}
\label{equation1}
S(\omega,t)=\left|\int_{-\infty}^{\infty}E(t')G(t-t')\mathrm{e}^{-i\omega t'}dt' \right|^2
\end{equation}
where $G(t-t')$ is a gate function, whose temporal delay $t$-$t'$ can be varied. In standard non-linear FROG techniques \cite{Trebino:book} an ultrashort laser pulse is used as the temporal gate, while a thin slice of the spectrum is filtered by the sum-frequency process on the non-linear crystal. 
In our approach, the gate function is replaced by a step-like function represented by the VO$_{2}$ film reflectivity variation, triggered by a $\sim$120 fs-800 nm pump pulse (see Fig. \ref{fig1}). The activation time $t$-$t'$ is delayed through the duration of the SC pulse simply varying the relative optical path between the pump and probe pulses. In this configuration, the time resolution of the measured spectrogram is given by the time needed to switch the VO$_2$ optical properties, which is shorter than the 120 fs pulse duration \cite{Cavalleri2004}. Since the VO$_{2}$ reflectivity variation extends over a broad frequency range,  the spectral resolution is determined by the width of each slice of the spectrum acquired by the detection system.  

\begin{figure*}[t]
\includegraphics[keepaspectratio, bb= 20 190 570 660, clip,width=0.8\textwidth]{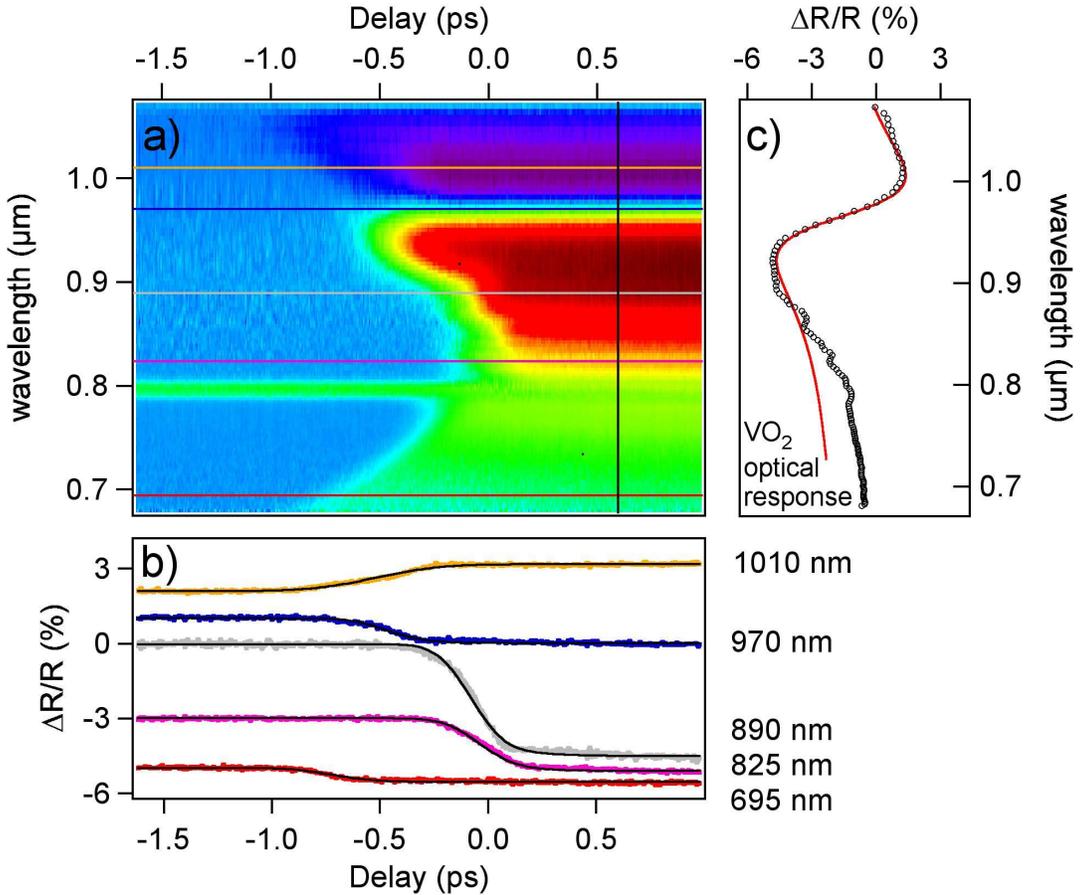}
\caption{\footnotesize{a) The time-and-frequency resolved reflectivity variation, measured on the VO$_{2}$ multi-film sample, is shown. b) Time traces at different wavelengths are reported. The solid lines are the fit to the data, as described in the text. c) The frequency-resolved reflectivity variation at $t$-$t'$=0.6 ps is reported. The solid line is the reflectivity variation mimicked replacing the VO$_{2}$ insulating dielectric function with metallic one.}}
\label{fig2}
\end{figure*}

The optical pump-probe setup is based on a cavity-dumped Ti:sapphire oscillator. The output is a train of  800 nm-120 fs pulses with an energy of 40 nJ/pulse.
The 15 cm-long PCF with a 1.6 $\mu$m core is pumped with 5 nJ/pulse focused into the core by an aspherical lens. The SC-probe output is collimated and re-focused on the sample through achromatic doublets in the near-IR/visible range. 
The 800 nm oscillator output (pump) and SC beam (probe), orthogonally polarized, are noncollinearly focused onto the sample, checking their superposition and dimensions (40 $\mu$m for the IR-pump and 20 $\mu$m for the SC probe) with a CCD camera. The pump fluence is about 3 mJ/cm$^{2}$, well above the VO$_{2}$ phase transition threshold \cite{Rini2008}.
The reflected probe beam is spectrally dispersed through a SF11 equilateral prism and imaged on a 128-pixel linear photodiode array (PDA), capturing the 700-1100 nm spectral region. A spectral slice, whose width ranges from 2 nm at 700 nm to 6 nm at 1100 nm, is acquired by each pixel of the array, corresponding to a constant temporal resolution of $\sim$120 fs.
The probe beam is sampled before the interaction with the pump and used as a reference for the SC intensity.
The outputs of the two PDAs are acquired through a 22-bit/2-MHz fast digitizer, and are divided pixel by pixel to compensate the SC intensity fluctuations,
obtaining a signal to noise ratio of the order of 10$^{-4}$ acquiring 2000 spectra (\begin{math} \sim \end{math}1 s integration time, cfr \cite{Polli2007}).
The differential reflectivity signal is obtained modulating the pump beam with a mechanical chopper and performing the difference between unpumped and pumped spectra. The multilayer sample is a 20 nm VO$_{2}$ thin film on a 330 nm Si$_{3}$N$_{4}$ buffer layer, deposited on a silicon substrate.

In Fig. \ref{fig2}a we report the frequency resolved VO$_{2}$ reflectivity variation, measured as a function of the delay between the supercontinuum pulse and the VO$_{2}$ switching, triggered by the pump pulse. A decrease of the reflectivity of a few percent is measured below 970 nm, whereas, above this value, the reflectivity variation becomes positive. The onset of the reflectivity variation exhibits a parabolic shape with wavelength, related to the parabolic group delay variation with wavelength typical of PCF-generated supercontinuum pulses \cite{Dudley2002}. 
The flat signal measured at 800 nm is due to a low intensity post-pulse produced by the cavity and entering in the PCF about 18 ps after the main pulse. The small intensity ($<$1$\%$ of the main pulse) of the post-pulse prevents it from non-linearly interacting with the fiber and broadening its spectrum. We underline that the switching character of VO$_{2}$ reveals the presence of post-pulses even if the investigated temporal window is small as compared to the delay between the main and the secondary pulses. In Fig. \ref{fig2}b different time-traces at specific wavelengths are shown. A convolution between a step function centered at $t$-$t'$, representing the VO$_{2}$ reflectivity variation, and a gaussian is fitted to the time-traces (solid lines in Fig. \ref{fig2}b). The gaussian takes into account both the temporal resolution (pulse width and VO$_{2}$ response) and the time width of the wavepacket centered at the frequency $\omega$ detected by the pixel.  The extracted $t$-$t'$ values and gaussian widths represents the delays and the temporal widths, respectively, of the particular slices of the spectrum corresponding to different pixels of the detector.

To give more insights into the origin of the ultrafast variation of the VO$_2$ optical properties, in Fig. \ref{fig2}c we report the frequency-resolved reflectivity variation, measured at a delay of +0.6 ps from the pump pulse and representing the optical response of the device. The complex reflectivity variation exhibited by the VO$_{2}$ sample is the result of both the occurrence of the insulator-to-metal phase transition and the multiple internal reflections at the multi-film interfaces. The measured reflectivity variation has been qualitatively reproduced (solid line in Fig. \ref{fig2}c), taking into account the sample geometry and replacing the dielectric constant of the VO$_{2}$ insulating phase with that of the metallic phase. In particular, the insulating dielectric function is described by four Lorentz oscillators with frequencies at 1.25 eV, 2.8 eV, 3.6 eV and 4.9 eV \cite{Verleur1968}, whereas the optical properties of the conducting phase are mimicked substituting the 1.25 eV band-gap related oscillator with a Drude term, representing the metallic free electrons, and slightly varying the energies, plasma frequencies and damping of the other oscillators, to take into account the shift of the O 2$p$ and V 3$d$ bands, related to the change in the crystal structure \cite{Cavalleri2005}. The refraction indexes of the other materials constituting the multi-film sample have been taken from Ref. \onlinecite{Luxpop}.
The fact that the photo-induced variation of the sample optical response extends throughout a broad spectral range, confirms the choice of this system as an ultrafast optical switch. 

\begin{figure*}[t]
\centering
\includegraphics[keepaspectratio, bb= 20 290 580 540, clip,width=0.8\textwidth]{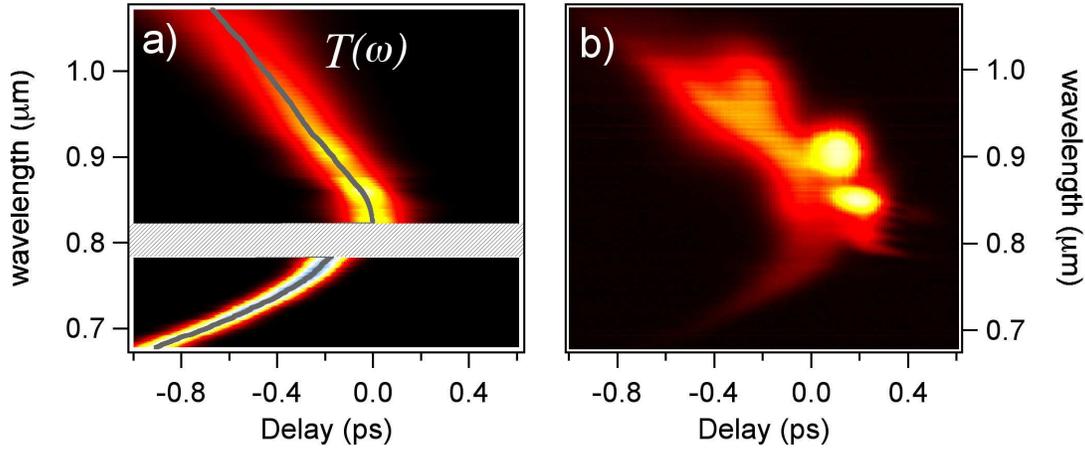}
\caption{\footnotesize{a) The $T(\omega)$ of the SC pulse, reconstructed through the VO$_{2}$-based technique described in the text, is reported. The solid line is the $t$-$t'$ delay between the step function, mimicking the VO$_{2}$ switch opening, and the probe pulse. b) The SC spectrogram, measured through a standard XFROG technique, is shown.}}
\label{fig3}
\end{figure*}

In Fig. \ref{fig3} we compare the $T(\omega)$ of the SC pulse (Fig. \ref{fig3}a), reconstructed through the procedure described above, to the trace (Fig. \ref{fig3}b) obtained through a standard cross-correlation FROG technique (XFROG) \cite{Dudley2002}. The XFROG trace has been obtained performing a cross-correlation between the 800 nm pump pulse and the SC probe pulse and recording the sum frequency (SF) intensity, generated in the angle-dithered BBO crystal, through a UV spectrometer. While the VO$_{2}$-based method exhibits a poorer time-resolution, as compared to the XFROG technique, all the main temporal features of the SC pulse can be reconstructed. In particular, the parabolic group delay variation with wavelength of the PCF-generated pulse can be satisfactorily measured. The limitations in the time resolution are mainly related to the pump pulse time-width, to the fitting procedures and to the velocity of the impulsive variation of the VO$_{2}$ optical properties, which is limited, in the present case, by the sample quality and by the intrinsic finite time for the  VO$_{2}$ switching. Improving the quality of the device and optimizing the fitting procedures will appreciably increase the resolution of this new solid-state based technique. Alternatively, different materials with a more rapid switching time should be considered \cite{Polli2007b}.

Finally, the VO$_{2}$-based technique exhibits some fundamental advantages as compared to the XFROG: i) the SC pulse characterization can be performed in the reflectivity configuration, without the need of phase-matching conditions. This permits to characterize SC pulses in situ, close to samples where time-and-frequency reflectivity experiments are performed, without the need to replace the sample holder with an angle-dithered BBO crystal. ii) Multiple-pulses, severely affecting ultrafast laser applications, can be easily detected without the need of time-consuming scanning of the entire temporal window between two subsequent pulses. \\

\begin{acknowledgements}
We acknowledge the comments from C. Manzoni.
\end{acknowledgements}


\end{document}